\documentclass[options,onecolumn]{mn2e}
\usepackage{graphicx}
\usepackage{subfigure}

\newif\ifAMStwofonts

\title[The integrated bispectrum as a test of CMB non-Gaussianity]
{The integrated bispectrum as a test of CMB non-Gaussianity: detection 
power and limits on $f_{NL}$ with WMAP data} 

\author[P. Cabella, F.~K. Hansen, M. Liguori, D. Marinucci, S. Matarrese, 
L. Moscardini, N. Vittorio]
{P. Cabella$^1$\thanks{E-mail: paolo.cabella@roma2.infn.it},
F.~K. Hansen$^2$\thanks{E-mail: f.k.hansen@astro.uio.no}, 
M. Liguori$^{3,4,5}$\thanks{E-mail: michele.liguori@pd.infn.it}, 
D. Marinucci$^6$\thanks{E-mail: marinucc@mat.uniroma2.it}, 
S. Matarrese$^{4,5}$\thanks{E-mail: sabino.matarrese@pd.infn.it},
\newauthor L. Moscardini$^7$\thanks{E-mail: lauro.moscardini@unibo.it} 
and N. Vittorio$^{8,9}$\thanks{E-mail: nicola.vittorio@roma2.infn.it}\\
$1$ Astrophysics, University of Oxford, Denys Wilkinson Building, Keble Road,
Oxford OX1 3RH, UK\\
$2$ Institute of Theoretical Astrophysics, University of Oslo, P.O. Box 1029 
Blindern, 0315 Oslo, Norway\\
$3$ Particle Astrophysics Center, Fermi National Accelerator Laboratory, 
Batavia, Illinois 60510-0500, USA\\
$4$ Dipartimento di Fisica 'Galileo Galilei', Universit\`a di Padova, 
Via Marzolo 8, I-35131 Padova, Italy\\
$5$ INFN, Sezione di Padova, Via Marzolo 8,  I-35131 Padova, Italy\\
$6$ Dipartimento di Matematica, Universit\`a di Roma `Tor Vergata', Via
della Ricerca Scientifica 1, I-00133 Roma, Italy  \\
$7$ Dipartimento di Astronomia, Universit\`a di Bologna, 
Via Ranzani 1,I-40127 Bologna, Italy\\
$8$ Dipartimento di Fisica, Universit\`a di Roma `Tor Vergata', Via
della Ricerca Scientifica 1, I-00133 Roma, Italy\\
$9$ INFN, Sezione di Roma `Tor Vergata',
 Via della Ricerca Scientifica 1, I-00133 Roma, Italy}

\pagerange{\pageref{firstpage}--\pageref{lastpage}}
\pubyear{2005}

\newcommand{\VEV}[1]{\langle#1\rangle}

\begin{document}

\label{firstpage}

\maketitle

\begin{abstract}
We propose a fast and efficient bispectrum statistic for 
Cosmic Microwave Background (CMB) temperature anisotropies 
to constrain the amplitude of the primordial non-Gaussian signal measured 
in terms of the non-linear coupling parameter $f_\mathrm{NL}$. 
We show how the method can achieve a remarkable computational advantage 
by focussing on subsets of the multipole configurations, where the 
non-Gaussian signal is more concentrated. The detection power of the test, 
increases roughly linearly with the maximum multipole, as shown in the 
ideal case of an experiment without noise and gaps. 
The CPU-time scales as $\ell_\mathrm{max}^3$ instead of $\ell_\mathrm{max}^5$ 
for the full bispectrum which for Planck resolution 
$\ell_\mathrm{max}\sim3000$ means an improvement in speed of a factor 
$10^7$ compared to the full bispectrum analysis with minor loss in 
precision. We find that the introduction of a galactic cut partially 
destroys the optimality of the configuration, which will then need 
to be dealt with in the future. 
We find for an ideal experiment with $\ell_\mathrm{max}=2000$ that upper 
limits of $f_\mathrm{NL}<8$ can be obtained at $1\sigma$. For the case of the 
WMAP experiment, we would be able to put limits of $|f_\mathrm{NL}|<40$ 
if no galactic cut were present. Using the real data with galactic cut, 
we obtain an estimate of $-80<f_\mathrm{NL}<80$ and $-160<f_\mathrm{NL}<160$ 
at 1 and $2\sigma$ respectively.

\end{abstract}



\section{Introduction}

In recent years a number of papers have focussed on the statistical nature
of the fluctuations in the Cosmic Microwave Background radiation (CMB). 
In the standard inflationary scenario the quantum fluctuations of the 
inflaton scalar field follow a nearly Gaussian distribution, with small 
deviations arise by considering second-order terms of the equations
\cite{acqua,malda}: slow-roll conditions necessarily entail that 
deviations from Gaussianity is very low. Nonetheless, the 
subsequent gravitational evolution unavoidably enhances the primordial 
non-Gaussian signal up to the largest scales where the primordial 
seeds were produced, giving rise to a non-linearity parameter (see below)
$f_{\rm NL}\sim {\cal O}(1)$.        
The situation changes in the presence of a second scalar field during 
inflation; under these circumstances it has been shown by 
(Bartolo Matarrese and Riotto 2002), that 
non-Gaussianity can be transferred from the isocurvature to 
the adiabatic mode, leading to non-negligible values for $f_{\rm NL}$ 
(see also \cite{uzan,rigopoulos,seery}). 
Alternative scenarios for the production of the primordial seeds, 
such as the {\it curvaton}(see \cite{mollerach,enqvist,lyth,moroi,lyth2}) and 
the {\it inhomogeneous reheating} mechanisms (see Dvali, Gruzinov and Zaldarriaga 2004) may also lead 
to higher values of $f_{\rm NL}$. 
A general review for the primordial non-Gaussian scenarios can be found 
in (Bartolo et al. 2004). 
Testing for non-Gaussianity has then become the basic tool to discriminate 
among different models for the production of energy-density perturbations. 
It has become common practice to quantify the amount of non-Gaussianity 
by means of the dimensionless {\it non-linearity parameter} 
$f_\mathrm{NL}$ (see, {\it e.g.} \cite{ks}), 
setting the strength of quadratic non-linearities in an expansion 
of the large-scale gravitational potential $\Phi$ (conventionally 
defined so that the
temperature anisotropy is $\Delta T/T \equiv - \frac{1}{3}\Phi$ in the
Sachs-Wolfe limit) in terms of a Gaussian random field $\Phi_{\rm G}$, namely 
\begin{equation}
\Phi({\bf x}) = \Phi_{\rm G}({\bf x}) + f_{\rm NL} \Phi_{\rm G}^2({\bf x}) \;
\end{equation} 
(up to a constant offset, which only affects the monopole contribution). 
Detailed calculations of the non-linearity parameter $f_{\rm NL}$ during
and after inflation (see \cite{bmrPRL,bkmr04}) 
have shown that it unavoidably contains an angle-dependent part, whose 
role could be extremely important to look for specific signatures of 
inflationary non-Gaussianity as recently shown in \cite{michele2}. In what follows, 
however, we will follow the common practice of taking $f_{\rm NL}$ as a
constant parameter. 
The upper limits on the estimated value of $f_\mathrm{NL}$ have become 
more and more stringent 
as the sensitivity of CMB experiments has improved. With MAXIMA data, 
\cite{santos} put a limit of  $|f_\mathrm{NL}|< 950$, at 1 $\sigma$ level.
Using  COBE data, \cite{komatsua} (with the bispectrum) and 
\cite{cayona} (with Spherical Mexican Hat Wavelets (SMHW)) found 
$|f_\mathrm{NL}|<1500$ and $|f_\mathrm{NL}|< 1100$ respectively, 
at 1 $\sigma$ level; these intervals were shrinked to 
$-58<f_\mathrm{NL}<134$ at 2$\sigma$ level in the first release of 
WMAP see \cite{komatsu}. Using SMHW, \cite{wang} found 
$f_\mathrm{NL}=50\pm80$ at 1 sigma and $ f_\mathrm{NL}<220 $ 
at 2$ \sigma$ level; in \cite{cabella1} we constrained 
$f_\mathrm{NL}=-5\pm 175$ at 2$\sigma$ level, combining 
the local curvature and spherical wavelets. Very recently 
the constraints on $ f_\mathrm{NL}$ have been improved by 
\cite{creminelli}, who find $-27<f_\mathrm{NL}< 121$ at 2$\sigma$ 
level. In \cite{gaztanaga} very stringent limits are found but a 
direct comparison with the other methods is unfeasible because of 
discrepancies in the non-Gaussian models adopted.\\
\noindent In this paper we focus on functionals of the normalized 
bispectrum; the
latter has been considered by many authors in the literature, 
including \cite{ks}; more recently, \cite{Babich} has discussed 
conditions under which the bispectrum is the optimal
estimator of primordial non-Gaussianity and \cite{babzalda} 
showed that tighter constraints are expected from a joint 
analysis of temperature and polarization data.\\
\noindent Here, we implement some procedures 
which were proposed in the statistical literature \cite{marinucci}. In that
paper, a full analytic derivation is provided for the test behaviour in the
presence of an ideal experiment; the behaviour in the presence of 
non-Gaussianity is also discussed. Here, we investigate the properties 
of these tests under a realistic experimental setting, using both simulations
and WMAP\ data. The plan of this paper is as follows: in Section 
\ref{sect:test} we describe the proposed procedure; Section 
\ref{sect:impl} and \ref{sect:data} describe the implementation, 
simulations and datasets used; Section \ref{sect:MC} discusses 
the results of the procedure applied to the simulations; in Section 
\ref{sect:concl} we draw some conclusions and discuss directions for 
future research. 

\section{The integrated Bispectrum}

\label{sect:test} As well-known, the angular bispectrum $%
B_{l_{1}l_{2}l_{3}}^{m_{1}m_{2}m_{3}}$ is defined by%
\begin{equation}
B_{l_{1}l_{2}l_{3}}^{m_{1}m_{2}m_{3}}=
\langle a_{l_{1}m_{1}}a_{l_{2}m_{2}}a_{l_{3}m_{3}} \rangle
\label{vene2}
\end{equation}
As shown by \cite{Hu} for a statistically isotropic field it is convenient
to focus on the angle-averaged bispectrum, defined by 
\begin{equation}
B_{l_{1}l_{2}l_{3}}=\sum_{m_{1}=-l_{1}}^{l_{1}}\sum_{m_{2}=-l_{2}}^{l_{2}}%
\sum_{m_{3}=-l_{3}}^{l_{3}}\left( 
\begin{tabular}{lll}
$l_{1}$ & $l_{2}$ & $l_{3}$ \\ 
$m_{1}$ & $m_{2}$ & $m_{3}$%
\end{tabular}%
\ \right) B_{l_{1}l_{2}l_{3}}^{m_{1}m_{2}m_{3}};  \label{parr}
\end{equation}%
the minimum mean square error estimator is provided by%
\[
\widehat{B}_{l_{1}l_{2}l_{3}}=\sum_{m_{1}=-l_{1}}^{l_{1}}%
\sum_{m_{2}=-l_{2}}^{l_{2}}\sum_{m_{3}=-l_{3}}^{l_{3}}\left( 
\begin{tabular}{lll}
$l_{1}$ & $l_{2}$ & $l_{3}$ \\ 
$m_{1}$ & $m_{2}$ & $m_{3}$%
\end{tabular}%
\ \right) (a_{l_{1}m_{1}}a_{l_{2}m_{2}}a_{l_{3}m_{3}})\mathrm{.} 
\]

The distribution of the previous statistic depends on the angular
power spectrum of the CMB. It is a standard practice to make the angular
bispectrum model independent (under Gaussianity) by focussing on the
normalized bispectrum, which we define by
\begin{equation}
I_{l_{1}l_{2}l_{3}}=(-1)^{(l_{1}+l_{2}+l_{3})/2}\frac{\widehat{B}
_{l_{1}l_{2}l_{3}}}{\sqrt{C_{l_{1}}C_{l_{2}}C_{l_{3}}}}\mathrm{\ .}
\label{ult1}
\end{equation}%
The factor $(-1)^{(l_{1}+l_{2}+l_{3})/2}$ is usually not included in the
definition of the normalized bispectrum; it corresponds, however, to the
sign of the Wigner's coefficients for $m_{1}=m_{2}=m_{3}=0$, and thus it
seems natural to include it to ensure that $I_{l_{1}l_{2}l_{3}}$ and $%
b_{l_{1}l_{2}l_{3}}$ share the same parity (see (\ref{parr})).
An alternative estimator of $I_{l_{1}l_{2}l_{3}}$, which uses the
estimated rather than the theoretical bispectrum, is provided by
\[
\widehat{I}_{l_{1}l_{2}l_{3}}=(-1)^{(l_{1}+l_{2}+l_{3})/2}\frac{\widehat{B}%
_{l_{1}l_{2}l_{3}}}{\sqrt{\widehat{C}_{l_{1}}\widehat{C}_{l_{2}}\widehat{C}%
_{l_{3}}}}\mathrm{,} 
\]%

\[
\widehat{C}_{l}=\frac{1}{2l+1}\sum_{m=-l}^{l}|a_{lm}|^{2} 
\]
is the power spectrum of the given realization. In \cite{marinucci} has shown that, 
under Gaussian hypothesis, the two normalizations are equivalent.  
A crucial issue relates to how one can combine the information 
from the different multipoles into a single statistic. For 
statistically isotropic
fields the bispectrum can be non-zero only for configurations where $%
l_{1}+l_{2}+l_{3}$ is even and the triangle conditions hold, $%
|l_{i}-l_{j}|\leq l_{k}\leq l_{i}+l_{j},$ $i,j,k=1,2,3.$ It is not difficult
to see that if we avoid repetitions there are asymptotically $L^{3}/24$ such
configurations, where $L$ denotes the highest observable multipole; it is
therefore computationally very hard to consider the full set of bispectrum
ordinates for high resolution experiments such as WMAP or Planck. Various
solutions have been considered, see for instance \cite{komat2}. In
this paper, our idea is to restrict the analysis to a subset of the
bispectrum ordinates where the bulk of information on non-Gaussianity is
condensed.

The procedure we shall consider has been advocated in the statistical
literature by \cite{marinucci}: More precisely, for finite integers $%
l_{0}\geq 2,$ $K\geq 0$ we shall consider the processes:
\begin{equation}
J_{L;l_{0},K}(r)=\frac{1}{\sqrt{L}}\sum_{l=l_{0}+K+1}^{[Lr]-l_{0}-K}\left\{ 
\frac{1}{\sqrt{K+1}}\sum_{u=0}^{K}\widehat{I}_{l_{0}+u,l,l+l_{0}+u}\right\}
\label{eq:pas4}
\end{equation}%
where $[.]$ denotes the integer part of a real number; $0\leq r\leq 1$ and $%
l_{0}$ is an (arbitrary but fixed) value which can be taken equal to 2 or 3,
according to whether we wish to keep the quadrupole or not in the data. As
usual, the sums are taken to be equal to zero when the index set is empty. $%
K $ is a fixed pooling parameter: for $K=0$ we obtain the special case%
\begin{equation}
J_{L;l_{0}}(r)=\frac{1}{\sqrt{L}}\sum_{l=l_{0}+1}^{[Lr]-l_{0}}\widehat{I}%
_{l_{0},l,l+l_{0}}.  \label{pas2}
\end{equation}
The normalizing factors are chosen to ensure an asymptotic unit 
variance for all summands. In words, the strategy for $J_{L;l_{0},K}(r)$ 
is to look at
collapsed configurations; more precisely, for a fixed $l_{0}$ we aim at
maximizing the distance among multipoles, albeit preserving the triangle
conditions $l_{i}\leq l_{j}+l_{k}.$ For an ideal experiment, it is shown in
\cite{marinucci} that, as $L\rightarrow \infty ,$ for any fixed integers $%
l_{0}>0,$ $K\geq 0$
\begin{equation}
J_{L;l_{0},K}(r)\Rightarrow W(r)\mathrm{,}0\leq r\leq 1\mathrm{,}  
\label{teo5.3}
\end{equation}
where $\Rightarrow $ denotes weak convergence and $W(r)$ standard Brownian
motion, that is, the Gaussian process with zero mean, independent increments
and variance $<W(r)^{2}>=r.$ The concept of weak convergence ensures that
the asymptotic distribution can be immediately derived for any continuous
functional of $J_{L;l_{0},K}(r);$ for instance 
\[
\Pr \left\{ \max_{0\leq r\leq 1}J_{iL;l_{0},K}(r)\geq x\right\} =2\Phi (-x)
 \:\:\mathrm{for\: all}\mathrm{\ }x\geq 0\mathrm{.,} 
\]
where  Pr stands for the probability and $\Phi $(.) denotes the cumulative 
distribution function of a
standard Gaussian variate; these values are well-known and tabulated, and
can be used to double-check the validity of Monte Carlo simulations.

We wish now to discuss the expected power of this procedure under simplified
circumstances: we shall work in the framework of an ideal experiment and a
pure Sachs-Wolfe model. The Monte Carlo evidence presented in the next
section suggests, however, that our conclusions have a much more general
validity. From (\ref{teo5.3}) we know that, approximately 
\[
Var\left\{ J_{iL;l_{0},K}(r)\right\} \sim r\mathrm{,for \:large}L ; 
\]%
also, it is well-known that the Sachs-Wolfe bispectrum can be approximated by
\begin{equation}
B_{l_{1}l_{2}l_{3}}=Gf_\mathrm{NL}h_{l_{1}l_{2}l_{3}}\left( 
\begin{array}{ccc}
l_{1} & l_{2} & l_{3} \\ 
0 & 0 & 0
\end{array}
\right) \left\{
C_{l_{1}}C_{l_{2}}+C_{l_{2}}C_{l_{3}}+C_{l_{1}}C_{l_{3}}\right\} \mathrm{,}
\label{sw2}
\end{equation}
where $G$ is a positive constant,
\[
h_{l_{1}l_{2}l_{3}}=\left( \frac{(2l_{1}+1)(2l_{2}+1)(2l_{3}+1)}{4\pi }%
\right) ^{1/2}, 
\]%
and lower order terms are neglected. We shall take $C_{l}\propto l^{-\alpha
} $ (for some positive constant $\alpha >2$). For simplicity, let us assume
that the normalizing angular power spectrum is known a priori; without loss
of generality, we take $K=0$. Using expressions 8.1.2.12 and 8.5.2.32 in
\cite{vals}, it can be shown that, for fixed $l_{0}\geq 2,$%
\[
\left( 
\begin{array}{ccc}
l_{0} & l & l+l_{0} \\ 
0 & 0 & 0%
\end{array}%
\right) =C\frac{(-1)^{l_{0}+l}}{\sqrt{l}}+O(\frac{1}{l^{3/2}})
\mathrm{,} 
\]%
for some $C>0$ which depends on $l_{0}$ but not on $l.$ Then we have easily
that%
\begin{equation}
<J_{L}(r)>\propto \frac{f_\mathrm{NL}}{\sqrt{L}}
\sum_{l=l_{0}+1}^{[Lr]}\sqrt{l}\sqrt{%
\frac{C_{l_{0}}C_{l}}{C_{l+l_{0}}}}\propto f_\mathrm{NL}L\mathrm{.}
\label{lig1}
\end{equation}
On the other hand, by a similar argument, it is easy to show that the choice
of an equilateral configuration $l_{1}=l_{2}=l_{3}=l$ entails an asymptotic
negligible power in the presence of this kind of non-Gaussianity.

The main conclusions we can draw from this heuristic discussion are as
follows: ``collapsed'' configurations where the multipoles lie on the
boundary of the triangle conditions seem to have an expected power of at
least an order of magnitude in $L$ larger than for configurations on the
main diagonal. For a pure Sachs-Wolfe model and an ideal experiment, the
signal to noise ratio is going to increase linearly for collapsed
configurations, whereas no improvement is expected for equilateral
configurations. It is then natural to conjecture that very little
information is lost in our procedure with respect to a full analysis of all
bispectrum coordinates; the latter, however, is clearly unfeasible for
computational reasons, and no analytic results are available to guide the
simulations. In fact whereas calculating all the elements of the 
bispectrum scales as $\ell^5$ in CPU time, calculating only the 
collapsed configurations scales as $\ell^3$. With WMAP data 
$(L\simeq 500)$ we gain a factor $10^5$ and for Planck 
resolution it is $10^7$ times faster. The next section is 
devoted to check the validity of the above 
claims in a much more general and realistic setting, by means 
of Monte Carlo simulations.

\section{The averaged integrated bispectrum of simulated non-Gaussian maps}

\label{sect:impl}
We have generated a set of 200 non-Gaussian simulations to test 
the power of the collapsed configurations described in the previous 
sections. Note that the above deductions were done for a model with 
fluctuations coming entirely from the Sachs-Wolf term with no radiative 
transfer involved. As the analytical treatment becomes much more complex 
with the radiative transfer function included, we will show that the 
above results are still valid using simulated non-Gaussian maps. First 
of all, we will show that for the 'Collapsed Configuration Bispectrum' 
(CCB) signal to noise is increasing for increasing multipoles, thus 
making the detection probability monotonically rising with increasing 
angular resolution. Second, we will show that the power of the CCB is 
falling with increasing $l_0$ and that all the power can be extracted 
using only the first few $l_0$. Finally, we will show that the often 
used diagonal configuration of the bispectrum has minimal power 
compared to CCB.

The non-Gaussian simulated maps were generated using the method of 
\cite{michele}. We produced a set of 200 maps (unfortunately, this is 
very CPU demanding limiting the number of maps which can be produced), 
the highest multipole being $L=2000$ with a power spectrum similar to 
the best fit WMAP power spectrum \cite{hinshaw}. The non-Gaussian part 
of the bispectrum can be obtain by making the ensemble average over 200 
simulations of
\[
\tilde J_{\ell}(L,L_0)\equiv\sum_{\ell_0=2}^{L_0}\frac{1}{\sqrt{L}}
\sum_{\ell'=\ell_0+1}^{\ell}\hat I_{\ell_0,\ell',\ell'+\ell_0}^\mathrm{NG},
\]
where $L$ is the maximum multipole used dependent on the resolution of 
the experiment and $L_0$ is the maximum value of $\ell_0$ included. Here 
NG means that we have only considered combinations of the type 
$a_{\ell_1 m_1}^\mathrm{NG}a_{\ell_2 m_2}^\mathrm{G}
a_{\ell_3 m_3}^\mathrm{G}$ in 
calculating the bispectrum. Since the non-Gaussian $a_{\ell m}^\mathrm{NG}$ 
are many orders of magnitude smaller than the corresponding gaussian 
$a_{\ell m}^\mathrm{G}$ combinations with more $a_{\ell m}^\mathrm{NG}$ 
factors will be negligible, also confirmed by our simulations. Note that 
in order to normalize the $a_{\ell m}$, we have assumed that we are able 
to estimate the power spectrum well and we use the correct ensemble 
averaged power spectrum. In the case of cut sky and noise, this is taken 
into account in the normalizing power spectrum by $C_\ell=C_\ell 
B_\ell^2/f_\mathrm{sky}+N_\ell$ where $B_\ell$ is the beam, $N_\ell$ 
is the noise power spectrum and $f_\mathrm{sky}$ is the sky fraction.

\begin{figure}
\includegraphics[width=15cm,height=10cm]{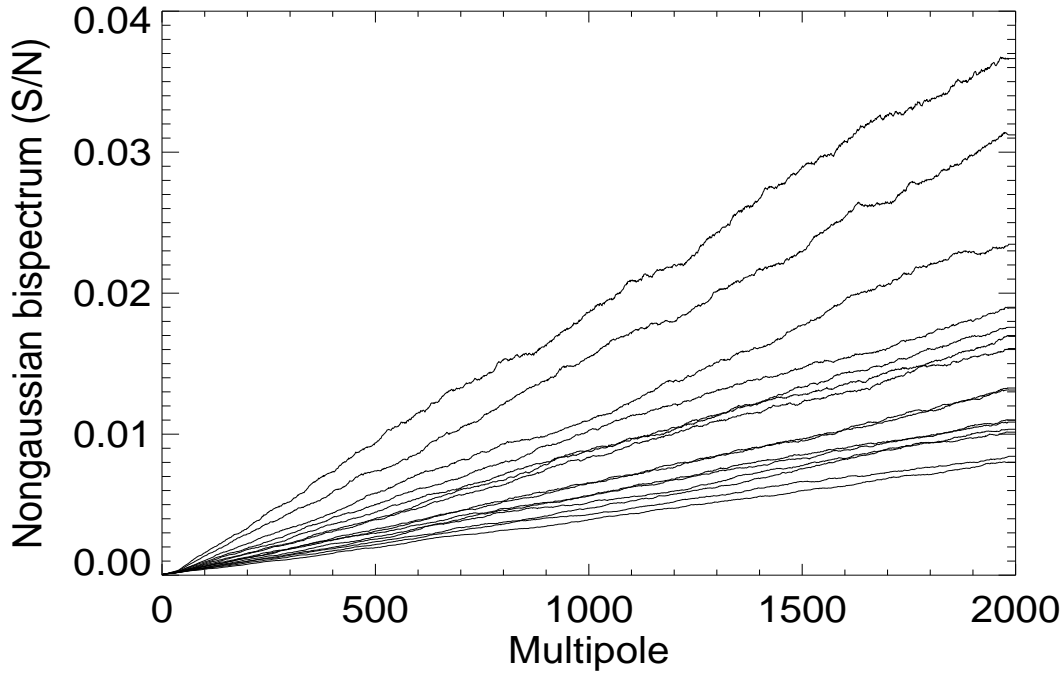}
\caption{The non-Gaussian part of the integrated bispectrum 
(see eq. \ref{eq:pas4}) averaged over 200 Monte Carlo simulations and 
normalized by the Gaussian standard deviation. Different values for 
$\ell_0$ (with constant step $\Delta \ell_0=1$) are shown, the lines are descending for increasing $\ell_0$, the highest being $\ell_0=2$, the lowest being $\ell_0=16$.} 
\label{fig:nongaussianCCB}
\end{figure}

In figure \ref{fig:nongaussianCCB}, we show the 
$\tilde J_{\ell, \ell_0}(L=2000)$ for different $\ell_0$. As expected, 
the non-Gaussian term is monotonically increasing and the contribution 
from high $\ell_0$ is decreasing. We also estimated the diagonal 
bispectrum from the same simulations and found it to be oscillating 
around 0, being at least three order of magnitude smaller than CCB.

\section{Simulated maps and the WMAP data}
 \label{sect:data}
In the following, we will demonstrate the power of CCB on simulated 
maps with noise and beam specifications being those of the WMAP 
experiment \cite{WMAP} (all data and templates are publicly available 
on the LAMBDA website\footnote{http://lambda.gsfc.nasa.gov/}). Finally, we will also analyse 
these data and compare our results to those of \cite{komatsu} in which 
all elements of the bispectrum are used. We will simulate the three CMB 
dominated WMAP channels, the Q(41 GHz), V(61 GHz) and W(94GHz) bands, 
convolving with the corresponding beams and adding Gaussian noise. We will 
also use the Kp0 galactic cut. Note that the galactic cut may shift the 
optimal configurations of the bispectrum and as we will show later, 
using a galactic cut seems to render the CCB less optimal. All analysis 
will be performed on the noise-weighted linear combination of the Q, V 
and W channels, co-added according to \cite{WMAP}
\[
T_i=(T_i^Qw_Q+T_i^Vw_V+T_i^Ww_W)/(w_Q+w_V+w_W),
\]
where $T_i^X$ is the temperature in pixel $i$ for channel X, and the 
weights are given as $w_X=1/\sigma^2_X$ where $\sigma^2_X$ is the 
average noise variance for channel X. We will simulate 200 Gaussian 
realizations of CMB with the corresponding 200 non-Gaussian maps  
\cite{michele} as well as 200 independent Gaussian simulations and noise 
for all 400 maps to be analysed.

\section{Estimating $\lowercase{f}_{\rm NL}$ in Monte Carlo simulations}

\label{sect:MC} 
     
The scope of this section is to find the error bars on $f_\mathrm{NL}$ 
for the simulated WMAP data described in the previous section. These 
error bars will be compared to the error bars obtained by \cite{komatsu} 
using all elements of the bispectrum as a test of the optimality of the 
CCB. We will apply the full estimation procedure to maps with and without 
galactic cut, checking in this way whether introducing a sky cut causes 
loss of optimality (thus that the optimal configurations are shifted away 
from the CCB by the non-ortonormality of the spherical harmonic functions 
on the cut sky). Finally we will estimate $f_\mathrm{NL}$ using the real data.

\begin{figure} 
\includegraphics[width=14cm,height=10cm, angle=0]{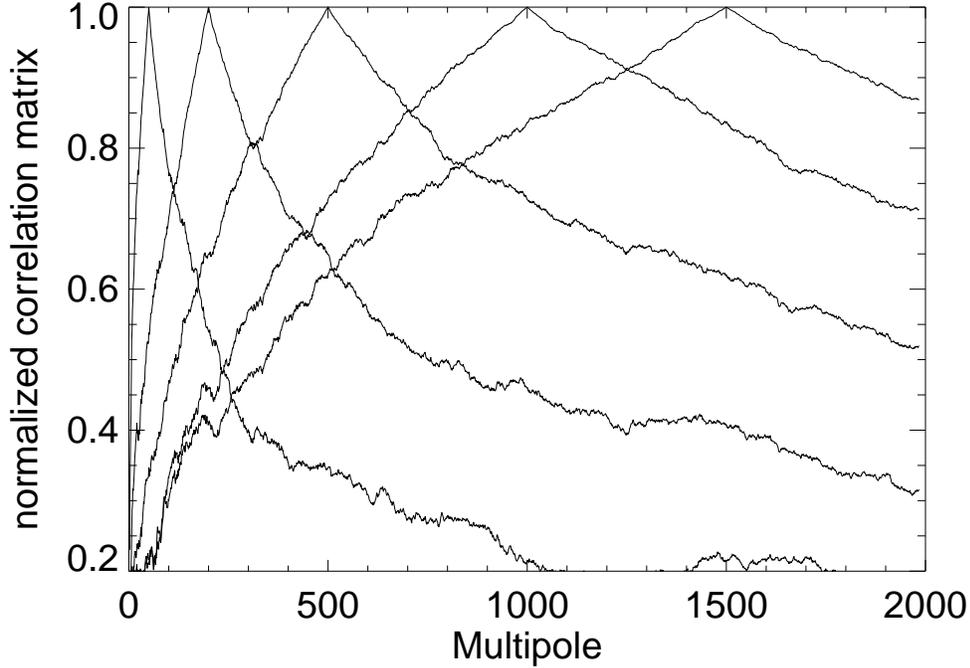}
\caption{Cross sections of the normalized correlation matrix. The plot 
shows $C_{\ell_0\ell'}/\sqrt{C_{\ell_0\ell_0}C_{\ell'\ell'}}$ for some 
values of ${\ell_0}$. The correlation matrix was obtained by 200 Gaussian 
simulations of the integrated bispectrum $C_{\ell\ell'}=\VEV{\tilde J_\ell 
\tilde J_{\ell'}}$.\label{fig:cormat} }
\end{figure}

In order to estimate $f_\mathrm{NL}$ from the CCB of the 200 simulated maps, 
we will minimize the $\chi^2$ defined by
\[
\chi^2(f_\mathrm{NL})={\bf d}(f_\mathrm{NL})^T{\bf C}^{-1}{\bf d}
(f_\mathrm{NL}),
\]
with respect to $f_\mathrm{NL}$. Here, the elements of the data vector 
are defined by $d_\ell=\tilde J_{\ell}-f_\mathrm{NL}*\VEV{\tilde 
J_{\ell}^\mathrm{NG}}$ and the correlation matrix $C_{\ell\ell'}=
\VEV{d_\ell d_{\ell'}}-\VEV{d_\ell}\VEV{d_{\ell'}}$ with $\tilde 
J_{\ell}$ being the 'observed' CCB. The 200 pure non-Gaussian maps 
are used to obtain $\VEV{J_{\ell}^\mathrm{NG}}$, and the 200 independent 
Gaussian simulations are used for obtaining the correlation matrix. 
Strictly speaking, the correlation matrix also depends on $f_\mathrm{NL}$ 
but for realistic values of $f_\mathrm{NL}$ ($<100$) this dependence 
is negligible. The normalized correlation matrix defined as 
$\hat C_{\ell\ell'}\equiv C_{\ell\ell'}/\sqrt{C_{\ell\ell}
C_{\ell'\ell'}}$ is shown in figure (\ref{fig:cormat}). The correlations
 between neighbouring multipoles is so strong that the matrix is 
numerically ill-defined. A binning procedure is necessary to enable 
the matrix to be inverted. Note from the figure that the long-range correlations are stronger for larger multipoles, 
thus a tighter binning for the lower multipoles will be allowed. We 
define the bin-size for a given multipole $\ell$ following this procedure:
\begin{itemize}
\item define a limit $\alpha<1$ 
\item start with $\ell=2$ and find for which $\ell$ the value of the 
normalized correlation matrix $\hat C_{2\ell}$ has fallen below the limit 
$\alpha$. This defines the next bin.
\item Starting with the obtained multipole $\ell'$ of the next bin, find 
for which multipole $\hat C_{\ell'\ell}$ has fallen below the limit $\alpha$.
\item repeat the above procedure until the highest multipole $L$ has been 
reached. Check if the correlation matrix with this binning gives a 
numerically well defined correlation matrix. If not, repeat the above 
procedure with a lower limit $\alpha$ otherwise, the binning procedure is 
finished and the final binning has been obtained.
\end{itemize}

\begin{figure}[]

\includegraphics[width=8cm,height=8cm, angle=0]{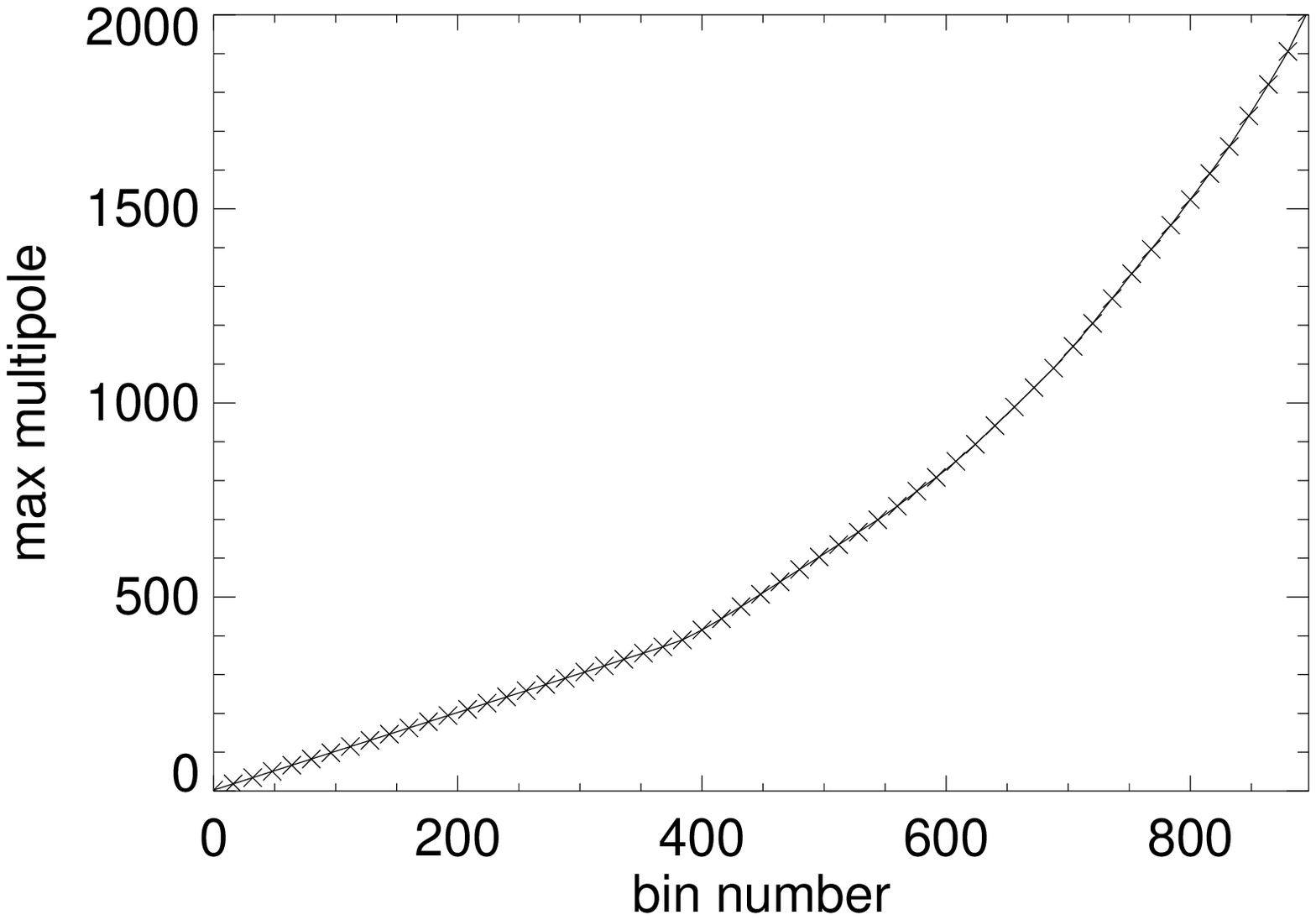}
\includegraphics[width=8cm,height=8cm, angle=0]{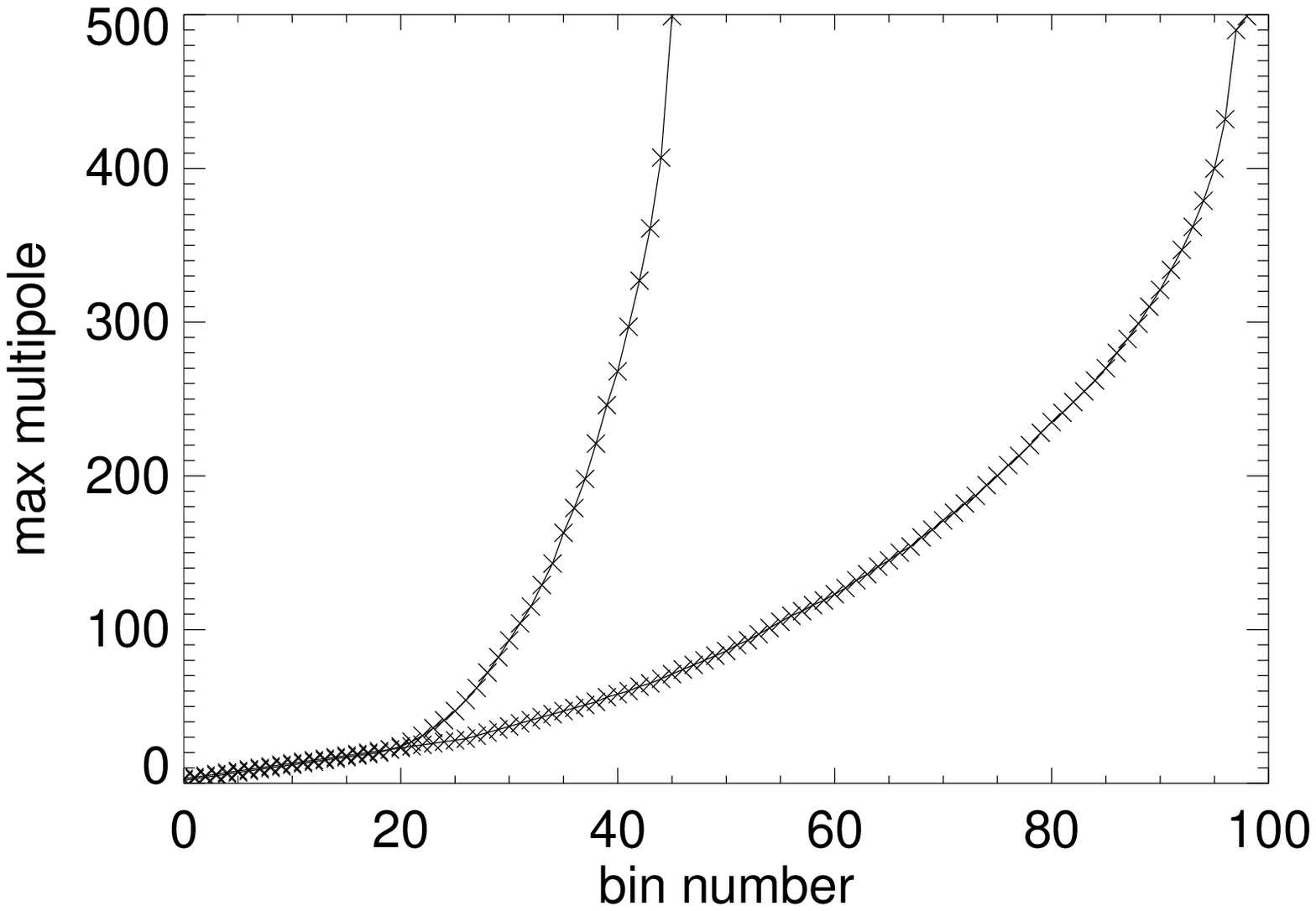}
\caption{Left panel: Multipole binning for the $\chi^2$ analysis of the high resolution 
simulations. On the x-axis we show the bin number and on the y-axis the 
corresponding last multipole of the bin. The crosses show each 16th bin 
to give an impression of the bin density per multipole. The limit 
$\alpha$ on the correlation matrix used to construct this binning (see 
the text) was $\alpha=0.998$. Right panel: the same, but for the WMAP simulations. The crosses are 
now shown for all bins. Lower line is for the case with WMAP noise but 
no galactic cut ($\alpha=0.985$, see the text), upper line is for the 
case with both noise and a Kp0 cut ($\alpha=0.946$) \label{fig:bin1}}
\end{figure}

Following this procedure, we obtained the binning shown in the panels of figure 
\ref{fig:bin1}, for the different cases which we will 
describe in the following. Further, we estimate $f_\mathrm{NL}$ in the 200 
Gaussian maps and obtain in this way the frequentist error bars. We have 
checked that the we obtain unbias estimates of $f_\mathrm{NL}$ and that 
error bars of maps with non-zero $f_\mathrm{NL}$ for realistic values 
($f_\mathrm{NL}<100$) do not deviate significantly from the value obtained 
on Gaussian realizations.

As a first test of this procedure, we estimated $f_\mathrm{NL}$ for an 
ideal experiment with no noise and no sky cut, $L=2000$ and $L_0=16$. 
In this case we obtained $\Delta f_\mathrm{NL}=8$ at $1\sigma$ being 
consistent with \cite{ks} who found $\Delta f_\mathrm{NL}=3$ for an 
ideal experiment with $L=3000$ using all elements of the bispectrum.

Second, we applied  the same procedure on the maps with WMAP beam and 
noise added (but no galactic cut) and find $\Delta f_\mathrm{NL}=40$ 
which is to be compared with $\Delta f_\mathrm{NL}=48$ obtained by 
\cite{komatsu} for the WMAP data. In the latter, a galactic cut of 
$25\%$ (Kp0) was applied. 

Finally, for the case including the Kp0 sky cut, we obtain 
$\Delta f_\mathrm{NL}=80$. Note that in this case the error has 
increased drastically and much more than one could expect from a 
simple $25\%$ loss of data. This can be understood, looking at the right panel of figure 
\ref{fig:bin1} showing the binning with and without the mask. The bins 
are much denser without the mask, showing that the mask is introducing 
high correlations between multipoles. Thus the information is spread to 
other multipoles and the collapsed configuration is no longer optimal. 
This suggests to include refilling procedure in order to restore 
orthogonality (i.e. one could use the Gibbs sampling approach 
\cite{eriksen2}) and just be limited by the sampling variance, 
this will be explored in a future work.         

\begin{figure}
\begin{center}
\includegraphics[width=8cm,height=8cm,angle=0]{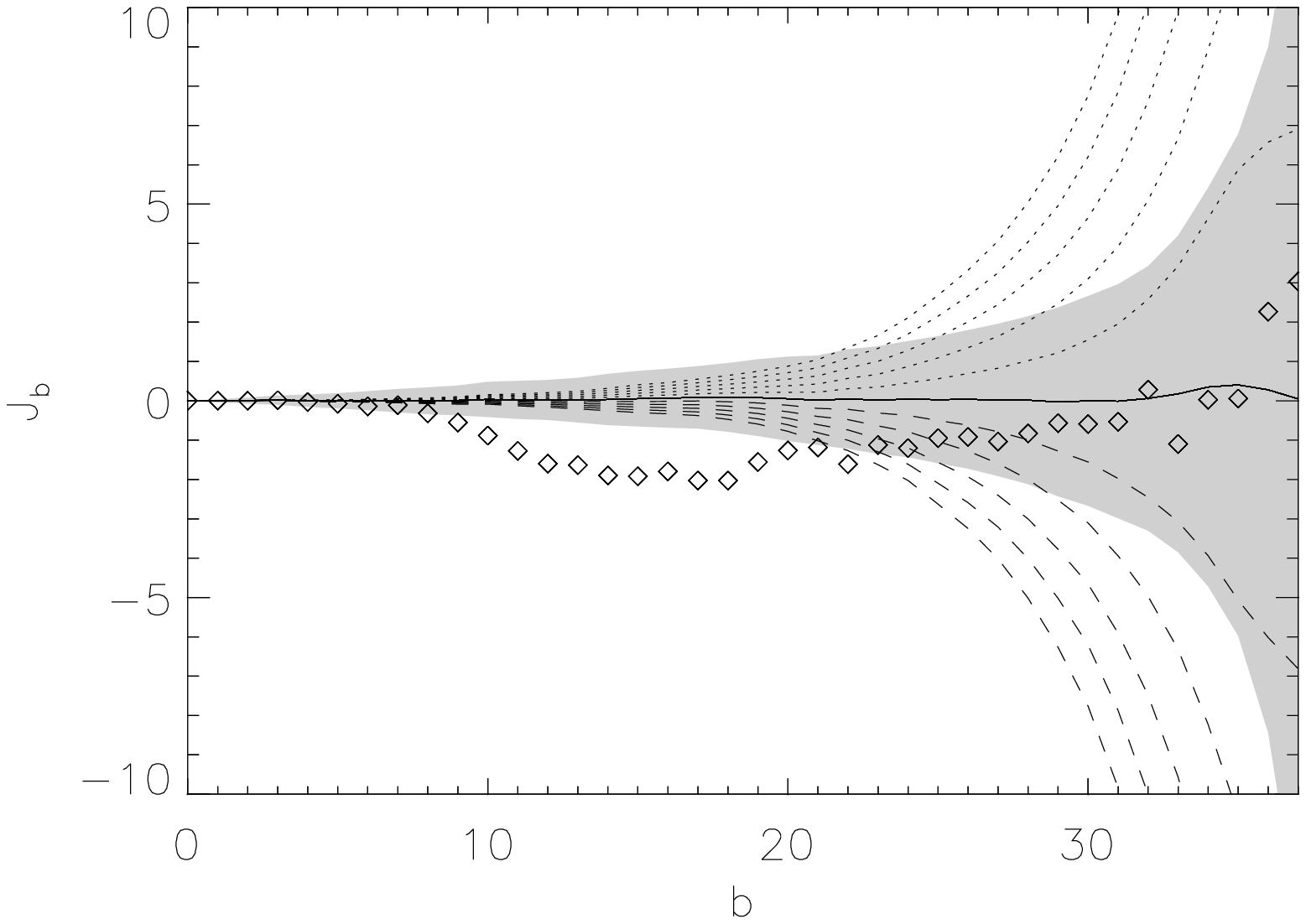}
\includegraphics[width=8cm,height=8cm,angle=0]{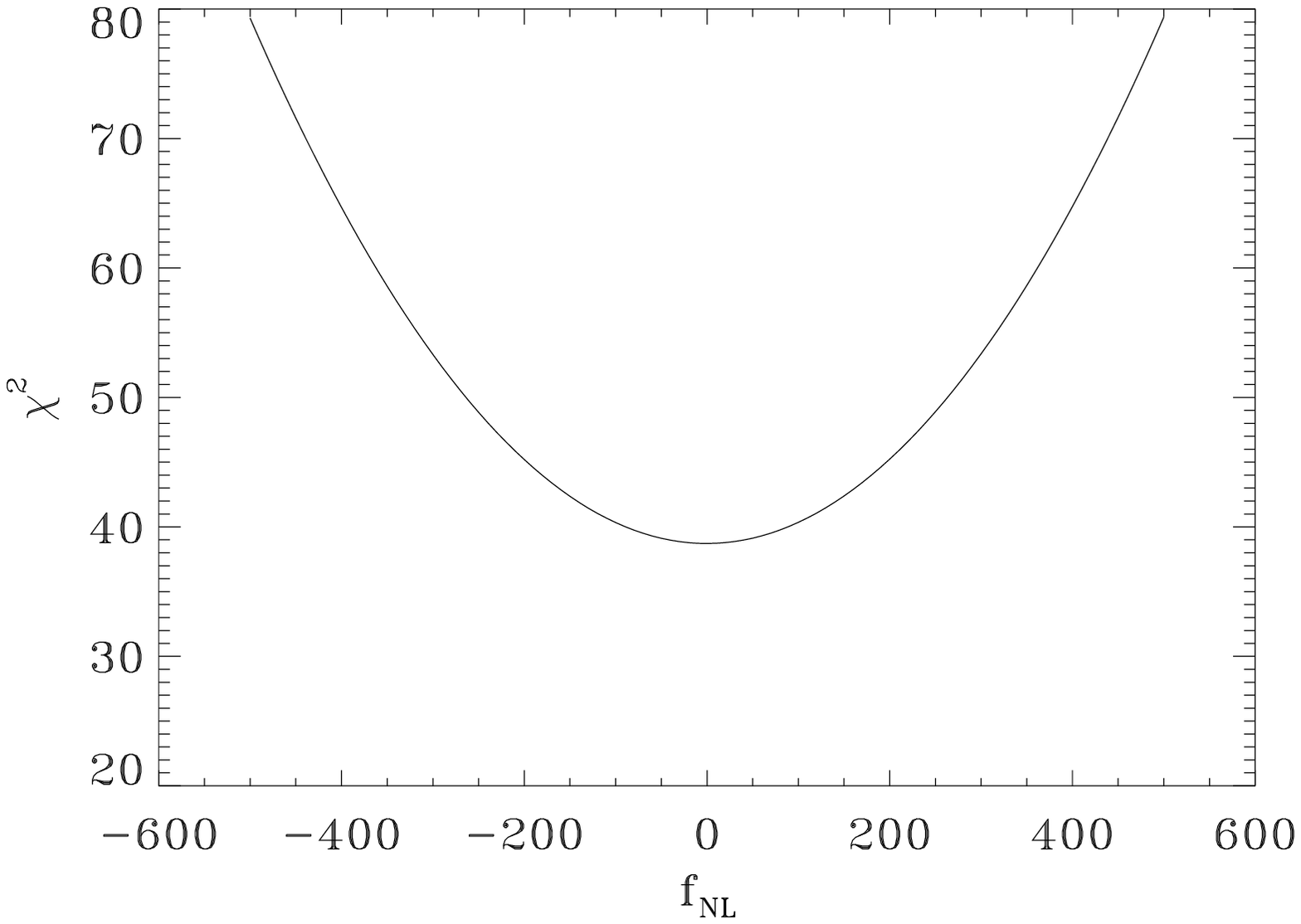}
\caption{Left panel: The integrated bispectrum (see eq. \ref{eq:pas4}) averaged
over 200 Monte Carlo simulations with WMAP noise and Kp0 mask. The solid 
line represents the mean of 200 Gaussian simulations, dotted (dashed) 
lines represent non-Gaussian simulations for different negative 
(positive) values  of $f_\mathrm{NL}$ from -500 to 500. The shaded area 
indicates the 1 $\sigma$ confidence level taken from the Gaussian 
simulations. The diamonds show the result of WMAP data. Right panel: The $\chi^2$ of WMAP data as a function of $f_\mathrm{NL}$. $f_\mathrm{NL}$ is estimated to be $0\pm 80$ and $0\pm 160$ at $1\sigma$ 
and  $2\sigma$ level respectively.\label{fig:MC} }
\end{center}
\end{figure}

We have applied the procedure to the WMAP data and obtain an estimate 
$f_\mathrm{NL}=0$ when sampling $f_\mathrm{NL}$ space in a grid of 10. 
In figure \ref{fig:MC}, we show the Gaussian standard deviation of 
$\tilde J_\ell$ together with $\tilde J_\ell$ for the WMAP data. We 
also show the theoretical $\tilde J_\ell$ for a set of $f_\mathrm{NL}$
 values. On the right panel of figure \ref{fig:MC}, we show the $\chi^2$ around its 
minimum, showing that the Bayesian error bars indicate error bars of 
$\Delta f_\mathrm{NL}$ about 80 at $1\sigma$ and 160 at $2\sigma$ in 
agreement with the frequentist analysis.

\section{Conclusions}

\label{sect:concl}

The bispectrum is one of the most common statistics 
to test for non-Gaussianity in CMB data. In particular, 
for estimating the non-linear coupling constant 
in primordial non-Gaussian models, the bispectrum has proven to produce very 
stringent limits. The drawback of the bispectrum is that the number of 
elements scale with the maximum multipole $L$ as $L^3$ and computational 
time scales as $L^5$.

In this paper we have introduced a new procedure to test for non-Gaussianity 
on CMB data; our approach is based on an integrated form of the bispectrum, 
where a phase factor is introduced and the focus is narrowed on 
nearly-collapsed configurations. Our approach is computationally very 
convenient as it scales only as $L^3$ and presents the added bonus to 
allow for explicit analytic results under idealized circumstances, in 
both Gaussian and non-Gaussian settings. The power properties are shown 
to be encouraging by Monte Carlo experiments: indeed some simple 
calculations suggest that the non-Gaussian signal grows linearly with 
the experiment resolution.

By comparing to limits on $f_\mathrm{NL}$ published in the literature 
 based on all elements of the bispectrum, we have shown that the
 collapsed configuration bispectrum does indeed seem to produce comparable 
near optimal results. For the ideal experiment we can constrain 
$f_\mathrm{NL}<8$ using $L=2000$ and for a WMAP like experiment we find 
$f_\mathrm{NL}<40$ when a galactic cut is not introduced. However, it 
turns out that a galactic cut does destroy the nice properties of the 
collapsed configuration. The multipoles get strongly coupled and the 
constrains on $f_\mathrm{NL}$ becomes bigger that what one would expect 
from a pure increase in sampling variance. For this reason, when the 
galactic cut is introduced, we obtain $-80<f_\mathrm{NL}<80$ at $1\sigma$ 
applied to the WMAP data. Clearly, data from future CMB experiments like 
Planck will still need to be analysed with a galactic cut as it will be 
impossible to completely eliminate the galactic plane. For that reason, a way of dealing with the non-optimality of the collapsed configuration 
bispectrum will be necessary. One possible solution to this would be by 
some refilling procedure (or one could use the already existing Gibbs 
sampling technique \cite{eriksen2}) which could restore orthogonality 
of the spherical harmonics. In this work, we did not make any prediction 
of the constrains on $f_\mathrm{NL}$ which can be achieved by the Planck 
experiment. As we are still unable to deal optimally with cut sky, this 
would not produce a precise limit and is therefore postponed to future work.

Finally, we note that for an ideal experiment it seems feasible to achieve 
the bounds predicted in \cite{ks}, despite the fact that we are using here 
only $L$, rather than $L^3$, bispectrum configurations. On the other hand, 
the presence of gaps greatly deteriorated the performance of our procedure. 
As the collapsed bispectrum seems potentially a very promising technique 
for high resolution data, we view the gap handling issue as a research 
priority in our future work.

\section*{Acknowledgements}

We acknowledge use of the HEALPix \cite{healpix} software and analysis
package for deriving the results in this paper. FKH was supported by a 
Marie Curie reintegration grant. We acknowledge the use of the Legacy 
Archive for Microwave Background Data Analysis (LAMBDA). Support for LAMBDA is
provided by the NASA Office of Space Science.

\end{document}